\documentclass{eas}
\usepackage{graphicx}
\newcommand{\nc}{\newcommand}

\newcommand{\CII}{[C\,{\sc ii}]}
\newcommand{\NII}{[N\,{\sc ii}]}
\newcommand{\NIII}{[N\,{\sc iii}]}

\newcommand{\OI}{[O\,{\sc i}]}

\newcommand{\HII}{H\,{\sc ii}}
\newcommand{\HI}{H\,{\sc i}}

\newcommand{\Halpha}{H$\alpha$}
\nc\micron{\mbox{$\mu$m}}
\nc{\cmcub}{\mbox{cm$^{-3}$}}
\nc{\cmsq}{\mbox{cm$^{-2}$}}
\nc{\Kkms}{\mbox{K~km~s$^{-1}$}}
\nc{\kms}{\mbox{km~s$^{-1}$}}
\nc{\mthirty}{\mbox{M\,33}}
\nc{\Tmb}{\mbox{$T_{\rm mb}$}}
\nc{\vlsr}{\mbox{v$_{\rm LSR}$}}
\nc{\twCO}{$^{12}$CO}
\nc{\thCO}{$^{13}$CO}
\nc{\msun}{\ensuremath{\mathrm{M}_\odot}}
\nc{\rsun}{\ensuremath{\mathrm{R}_\odot}}
\nc{\lsun}{\ensuremath{\mathrm{L}_\odot}}
\begin{document}

%%-----------------------------
%%      the top matter
%%-----------------------------
\title{Star formation in M33 (HerM33es)} 
\author{C.\,Kramer}\address{IRAM, Av. Divina Pastora 7, Nucleo Central, E-18012
  Granada, Spain } 
\author{M.\,Boquien}\address{Department of Astronomy, University of
Massachusetts, Amherst, MA 01003, USA}
\author{J.\,Braine}\address{Observatoire de Bordeaux, OASU, UMR 5804,
CNRS/INSU, Floirac F-33270}
\author{C.\,Buchbender$^1$}
\author{D.\,Calzetti$^2$}
\author{P.\,Gratier}\address{IRAM, 300 rue de la Piscine, 38406
St. Martin d'H\`{e}res, France}
\author{B.\,Mookerjea}\address{Tata Institute of Fundamental Research, Homi Bhabha
Road, Mumbai 400005, India}
\author{M.\,Rela\~{n}o}\address{Dept. F\'{i}sica Te\'{o}rica y del Cosmos,
Universidad de Granada, Spain}
\author{S.\,Verley$^6$}
%\author{M.\,Xilouris}\address{National Observatory of Athens, P. Penteli, 15236
%Athens, Greece}
%
%
\begin{abstract}
  Within the key project ``Herschel M33 extended survey'' ({\tt
    HerM33es}), we are studying the physical and chemical processes
  driving star formation and galactic evolution in the nearby galaxy
  M33, combining the study of local conditions affecting individual
  star formation with properties only becoming apparent on global
  scales. Here, we present recent results obtained by the {\tt
    HerM33es} team. Combining Spitzer and Herschel data ranging from
  3.6\,$\mu$m to 500$\,\mu$m, along with \HI, H$\alpha$, and GALEX UV
  data, we have studied the dust at high spatial resolutions of
  150\,pc, providing estimators of the total infrared (TIR) brightness
  and of the star formation rate. While the temperature of the warm
  dust at high brightness is driven by young massive stars, evolved
  stellar populations appear to drive the temperature of the cold
  dust.  Plane-parallel models of photon dominated regions (PDRs) fail
  to reproduce fully the \CII, \OI, and CO maps obtained in a first
  spectroscopic study of one $2'\times2'$ subregion of M33, located on
  the inner, northern spiral arm and encompassing the \HII\ region
  BCLMP\,302.
\end{abstract}
\maketitle
%%-----------------------------
%%      your text
%%-----------------------------
\begin{figure*}[t]   
  \centering   
  \includegraphics[width=10cm,angle=-90]{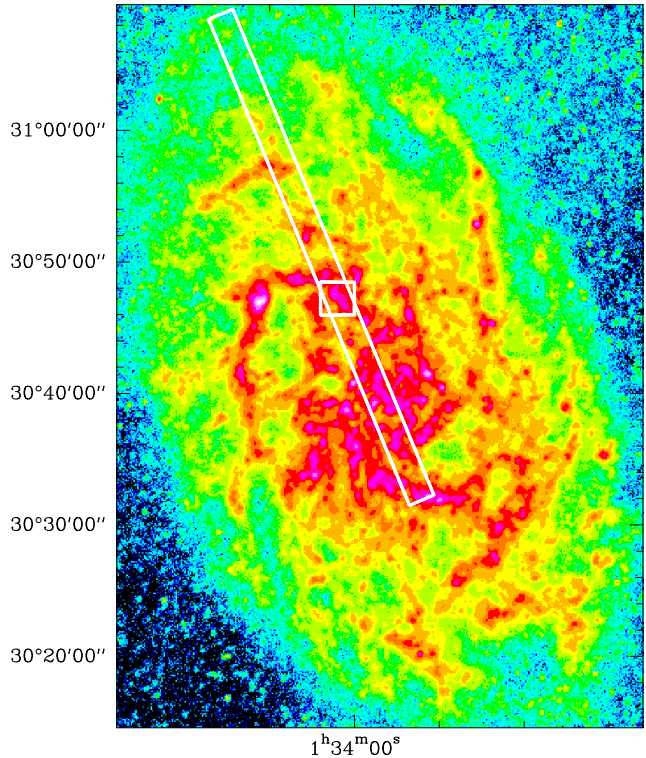} 
% from greg @ makeOverlays-m33-250tot.greg
%%%
  \caption{SPIRE 250\,$\mu$m map of M33 (Kramer et al. 2010).  The
    tilted rectangle delineates the $2'\times40'$ wide strip along the
    major axis, which will be partially mapped with HIFI and PACS in
    spectroscopy mode. The $2'\times2'$ box delineates the region
    mapped in \CII\ and \OI\ by Mookerjea et al. (2011) and shown in
    Figure 3.}
\label{fig-maps}   
\end{figure*}   

\section{Introduction}

%Within   the  key  project  ``Herschel   M33 extended  survey''  ({\tt
%  HerM33es}), we are studying    the physical and chemical   processes
%driving star formation and   galactic evolution in the   nearby galaxy
%M33, combining the study of local conditions affecting individual star
%formation with properties only becoming apparent on global scales.

M33 is a nearby galaxy located at only 840\,kpc with a moderate
inclination of $56^\circ$, actively forming stars. It has been studied
at radio, optical and X-ray wavelengths.  Its metallicity is about
half solar, similar to the LMC. Using Herschel, M33 provides an
exceptional physical resolution of $\sim50$\,pc at 158$\mu$m allowing
us to resolve the various morphological components of the galaxy. The
absence of an active nucleus excludes one possible dust heating
mechanism and the shallow metallicity gradient limits the influence of
the radial evolution of the metallicity on the emission of the dust.
In a first step to study the interstellar medium and its interplay
with star formation, we mapped the dust spectral energy distribution
over the entire galaxy using SPIRE and PACS (Kramer et al. 2010,
Verley et al.  2010, Braine et al.  2010, Boquien et al. 2010, 2011).

In a second step, we have started to use PACS and HIFI to map the
emission of \CII\ and other strong far-infrared (FIR) lines along the
major axis of M33 (Mookerjea et al. 2011).  Observing an extended
strip, will allow us to study the ionized, atomic, and molecular
phases of the interstellar medium, its life cycle and thermal balance,
tracing the formation of molecular clouds and of stars.

Here, we summarize the major results obtained in two recently
submitted papers of the {\tt HerM33es} team. But note that only
8\,hours out of the total granted 191\,hours of Herschel observing
time were observed till end of 2010.

\begin{figure*}[t]   
  \centering   
  \includegraphics[width=6cm]{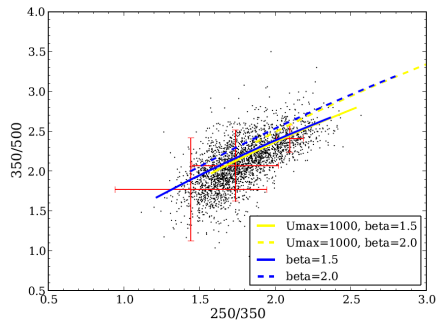} 
% from Boquien et al. 2011
%%%
  \caption{The observations of the SPIRE bands can be reproduced by
    the Draine \& Li (2007) models (Boquien et al. 2011). Here, we
    consider a single illuminating radiation field or a radiation
    field ranging from an arbitrary $U_{\rm min}$ to $U_{\rm
      max}$=1000, with $U$ the radiation field in the solar
    neighborhood. Even if the scatter is large, an emissivity index of
    $\beta=1.5$ fits the observations very well, in accordance with
    the results obtained from the radial averages studied by Kramer et
    al.  (2010).  }
\label{fig-maps}   
\end{figure*}   

\section{Dust heating sources in galaxies: the case of M33}

Dust emission in galaxies can have several unrelated origins: (1)
Massive stars in star-forming regions heat the dust. (2) A large scale
diffuse emission due to cirrus being heated by (a) energetic radiation
escaping from individual star-forming regions and/or (b) the general
radiation field of evolved stars. (3) Hot grains in the photosphere or
circumstellar atmosphere of mass-loosing stars also contribute. 

IR emission acts as a star-formation tracer even at high redshifts. As
the diffuse emission can represent up to nearly 90\% of the IR flux in
Sa galaxies, it can lead to an error on the SFR up to an order of
magnitude depending on its origin.

In Boquien et al. (2011), we have used Herschel PACS and SPIRE
observations in combination with Spitzer IRAC and MIPS data, GALEX FUV
data, as well as ground-based H$\alpha$ and HI maps of M33, in order
to characterize the warm and cold dust populations. Given the high
spatial resolution of even the longest wavelength bands ($\sim150\,$pc
at 500$\,\mu$m), we perform a pixel-to-pixel analysis of the data in
order to constrain the dust heating sources. To do so, we have
convolved all the data to the resolution of the SPIRE 500$\,\mu$m
band.  The maps have then been registered to common pixels of a size
of $42''$. For the analysis, only pixels with a signal-to-noise ratio
of at least 3 have been selected. The limiting factor is here the PACS
bands which are relatively shallow.  In the subsequent analysis at
least $\sim850$ pixels have been selected in each case.

We have studied the evolution of the dust colors both as a function of
the position in the galaxy (including the radial distance) and the
brightness.

To determine the TIR luminosities, we have fitted the models of Draine
\& Li (2007) to the SED of each spatial pixel from 8$\,\mu$m to
500\,$\mu$m and integrated the model from 1\,$\mu$m to 1\,mm. In
addition, we derived the star formation rate (SFR) from Calzetti et
al. (2007) combining H$\alpha$ and the 24\,$\mu$m emission. Plotting
the derived SFR against just one IR band, we find that the emission at
8\,$\mu$m
% and 24\,$\mu$m 
is sublinear, confirming previous results. The emission near the peak
of the emission between 70\,$\mu$m and 160\,$\mu$m is a linear
estimator of the SFR. Finally, the emission in SPIRE bands is an
increasingly super linear estimator of the SFR with an ever larger
scatter around the best fit.

In summary, we found the following results:

\begin{enumerate}
\item The colors of the warm and cold dust components seem to be
  predominantly driven by the evolution of the radiation field.
\item Combining any set of Spitzer and Herschel bands, we have
  provided correlations to estimate both the TIR brightness and the
  SFR, extending the results of Boquien et al. (2010) and Verley et
  al. (2010), which is of importance for the study of high redshift
  galaxies.
\item The color trends of the warm and the cold dust show that they
  are heated by different sources. At higher SFR, the warm dust
  temperature seems to be driven by star formation. As star formation
  weakens, the temperature is increasingly driven by another
  component, most likely the evolved stellar populations. The cold
  dust temperature seems to be driven by the old stellar population,
  with a tight correlation with the local stellar mass.
\end{enumerate}

\begin{figure*}[t]   
  \centering   
  \includegraphics[width=8cm]{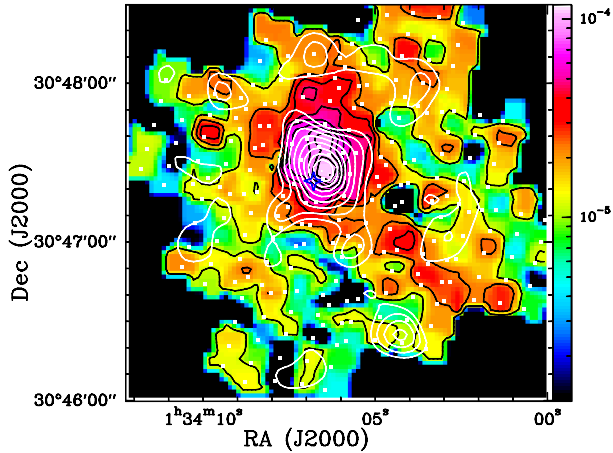}
% from Mookejea et al. 2011 
%%%
  \caption{Maps of 158\,$\mu$m\ \CII\ (in color and black contours)
    and 63\,$\mu$m\ \OI\ (white contours) emission observed with PACS
    toward the \HII\ region BCLMP\,302 in M33 (shown as small box in
    Figure\,1). The \Halpha\ peak observed with HIFI is marked with
    the asterisk (cf. Fig.\,4).  The white dots show the footprint of
    the PACS observations. Both images are at a common resolution of
    $12''$.}
\label{fig-pacs}   
\end{figure*}   

\begin{figure*}[h]   
  \centering   
  \includegraphics[width=6cm]{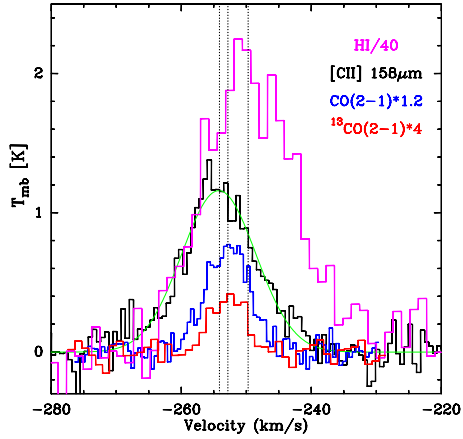}
% from Mookejea et al. 2011 
%%%
  \caption{Spectra of \CII, \HI, CO and \thCO\ 2--1 at the H$\alpha$
    peak position of the \HII\ region BCLMP\,302. All four spectra are
    at $\sim12''$ resolution.  }
\label{fig-spectra}   
\end{figure*}   

\begin{figure*}[h]
\centering
\includegraphics[width=6.0cm]{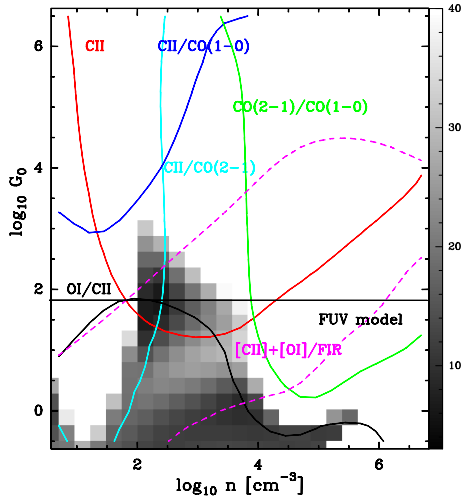}
\caption{Comparison of line intensities and intensity ratios at the
position of the \Halpha\ peak with plane-parallel constant density PDR
models (Kaufman et al. 1999) of different density $n$ and FUV field,
G$_0$. Grey-scales show the estimated reduced $\chi^2$. The horizontal
line shows the FUV estimated from the TIR intensity. 
\label{fig_hifipdrmod}}
\end{figure*}

\section{\CII\ and \OI\ at 50\,pc scales in the northern arm of M33}

In Mookerjea et al. (2011), we have used PACS on Herschel to observe
the emission of the far-infrared lines \CII\ (158\,$\mu$m), \OI\
(63\,$\mu$m), \NII\ (122\,$\mu$m), \NIII\ (57\,$\mu$m) in a
$2'\times2'$ region of the northern spiral arm of M\,33, centered on
the \HII\ region BCLMP\,302 (Figure\,3).

At the peak of H$\alpha$ emission, we have observed in addition a
velocity resolved \CII\ spectrum using HIFI (Figure\,4). The aim of
this work is to understand the relative contributions of the different
phases of the ISM to \CII\ cooling, as well as the correlation of the
\CII\ emission with the star formation rate (SFR), derived from
\Halpha\ and 24\,$\mu$m\ emission observed with Spitzer, at scales of
$12''$ corresponding to $\sim50$\,pc, i.e. at scales of individual
giant molecular clouds (GMCs).  We have obtained the distribution of
\CII\ and \OI\ (63\,$\mu$m) emission from the spiral arm and the
inter-arm regions, and detected the \NII(122\,$\mu$m) and
\NIII(57\,$\mu$m) lines at several positions.  
%
%The \NII(205\,$\mu$m) and \OI(145\,$\mu$m) lines were not detected.
These data are quantitatively compared with continuum maps observed
with PACS 100 \& 160\,$\mu$m\ and of CO and \HI\ data, at the same
resolution.

The \CII\ emission shows strong correlation with the TIR intensity
only within the \HII\ region and is well correlated with SFR along the
spiral arm. The gas heating efficiency, estimated as the ratio between
\CII\ and the TIR continuum, varies between 0.07 and 2\%. We used the
CLOUDY models of ionized and photon dominated regions (Ferland et al.
1998) and the known properties of BCLMP\,302, to estimate that upto
30\% of the \CII\ emission stems from the \HII\ region. Next, we used
plane-parallel constant density PDR models of Kaufman et al.  (1999),
to interpret the observed intensities of far-infrared and millimeter
lines at the H$\alpha$ position (Figure\,5).  Both, the FIR continuum
and results of PDR modeling indicate that a FUV field of
$\sim50$~G$_0$ is heating the dust and gas at the \Halpha\ peak
position. Over the surface of the studied region, the bulk of \CII\
emission originates from PDRs at cloud surfaces, while the
contributions from the atomic and molecular medium account for less
than 10\% of the observed \CII\ emission.

The rather poor fit of the intensity ratios towards the H$\alpha$ peak
shows the shortcomings of a plane-parallel single density PDR model.
First tests using KOSMA-$\tau$ PDR models (R{\"o}llig et al. 2006) of
spherical clumps with density gradients do better reconcile the
observations. In a forthcoming paper, we shall discuss new HIFI \CII\
data at other positions in the BCLMP\,302 region using much more
detailed PDR models, also taking into account the effects of sub-solar
metallicity.

%However, more sophisticated models considering a more complex density
%structure and geometry of the emitting regions, are probably needed to
%derive at a fully consistent interpretation of the emission.

%\section{...}
%\subsection{...}
%\subsection{...}
%
%%-----------------------------
%%      your bibliography
%%-----------------------------

\end{document}